\documentclass[
reprint,
nofootinbib,
amsmath,amssymb,
aps,
prx,
twocolumn,
floatfix,
]{revtex4-2}

\usepackage{amssymb}   
\usepackage{amsfonts}
\usepackage{amsmath}
\usepackage{mathtools}
\usepackage[mathscr]{eucal}

\usepackage{graphicx}
\usepackage{xcolor}

\usepackage{tikz}
\usetikzlibrary{quantikz2}

\DeclareUnicodeCharacter{FF0C}{}

\usepackage{hyperref}
\definecolor{bblue}{RGB}{0, 126, 204}
\definecolor{dgreen}{RGB}{0,102,51}
\definecolor{blueus}{RGB}{10, 49, 97}
\definecolor{teal}{RGB}{0,128,128}
\definecolor{orangered}{RGB}{255,69,0}
\hypersetup{
	colorlinks=true,       
	linkcolor=purple, 
	citecolor=bblue,        
	filecolor=magenta,      
	urlcolor=bblue           
}

\newcommand{\eq}[1]{\begin{align}#1\end{align}}

\begin{document}

\title{Simulation of a Rohksar-Kivelson ladder on a NISQ device}

\author{Sabhyata Gupta}
\email[]{sabhyata.gupta@itp.uni-hannover.de}

\author{Younes Javanmard}
\email[]{younes.javanmard@itp.uni-hannover.de}

\author{Tobias J. Osborne}
\email[]{tobias.osborne@itp.uni-hannover.de}

\author{Luis Santos}
\email[]{santos@itp.uni-hannover.de} 
\affiliation{Institut f\"ur Theoretische Physik, Leibniz Universit\"at Hannover, Appelstr. 2, 30167 Hannover, Germany}

\begin{abstract}
We present a quantum-classical algorithm to study the dynamics of the Rohksar-Kivelson plaquette ladder on NISQ devices. We show that complexity is largely reduced using gauge invariance, additional symmetries, and a crucial property associated to 
how plaquettes are blocked against ring-exchange in the ladder geometry. This allows for an efficient simulation of sizable plaquette ladders with a small number of qubits, well suited for the capabilities of present NISQ devices. We illustrate the procedure for ladders with simulation of up to $8$ plaquettes in an IBM-Q machine, employing scaled quantum gates.
\end{abstract}
\maketitle



\section{Introduction}
\label{sec:Introduction}


Recent years have witnessed spectacular progress in the field of quantum simulators~\cite{Georgescu2014, Altman2021}, i.e.\ quantum many-particle systems that simulate other quantum phenomena, which may not be tractable by classical means. This includes applications in disparate fields, ranging from condensed-matter physics~\cite{Bloch2012, Kennes2021} to material science and quantum chemistry~\cite{Cao2019,Bauer2020} and high-energy physics~\cite{Bauer2023, Bauer2023b}. Concerning the latter, quantum simulators open exciting possibilities for the simulation of quantum field theories, and in particular lattice gauge theories with both analogue and digital simulators~\cite{Wiese2013, Banuls2020,Aidelsburger2021,Bauer2023, Bauer2023b}. A particularly prominent example is provided by the successful simulation of the quantum link Schwinger model~\cite{Chandrasekharan1997}, a relatively simple model that describes quantum electrodynamics in one space and one time dimension~\cite{Hauke2013, Martinez2016,
 Klco2018, Kokail2019, Yang2020, Nguyen2022}. 
 
 
Despite these extraordinary developments, the analogue simulation of plaquette operators, crucial in the Hamiltonian formulation of lattice gauge theories~\cite{KogutSusskind},  remains very challenging, since they involve three- and higher-body interactions. Interestingly, it has been recently proposed that the Rydberg blockade in Rydberg configurable arrays could be employed to efficiently simulate plaquette terms, and in particular the Rohksar-Kivelson~(RK) model, a two-dimensional U(1) lattice gauge theory which has attracted a large deal of interest due to its relevance in the context of quantum dimer and spin ice theory~\cite{Moessner2011}.
 
 
 Digital simulations may overcome the limitations of analogue devices concerning plaquette operators~\cite{Lewis2019, Huffman2022}. However, at the present time, fully fault-tolerant quantum computers are not available. Instead, only noisy intermediate-scale quantum~(NISQ) devices~\cite{NISQPreskill2018quantumcomputingin} have been so far realised, with at most hundreds of qubits, and characterized by sparse connectivity and significant noise and decoherence in the application of quantum gates. Since quantum error correction is not yet possible, a set of error mitigation techniques~\cite{Cao2022,Cai2023}, and algorithms~\cite{Bharti2022} have been specifically designed for NISQ devices. In particular different variational quantum algorithms have been explored for the simulation of the Schwinger model~\cite{Klco2018, Avkhadiev2020}.



\begin{figure}[t!]
\includegraphics[width=\columnwidth]{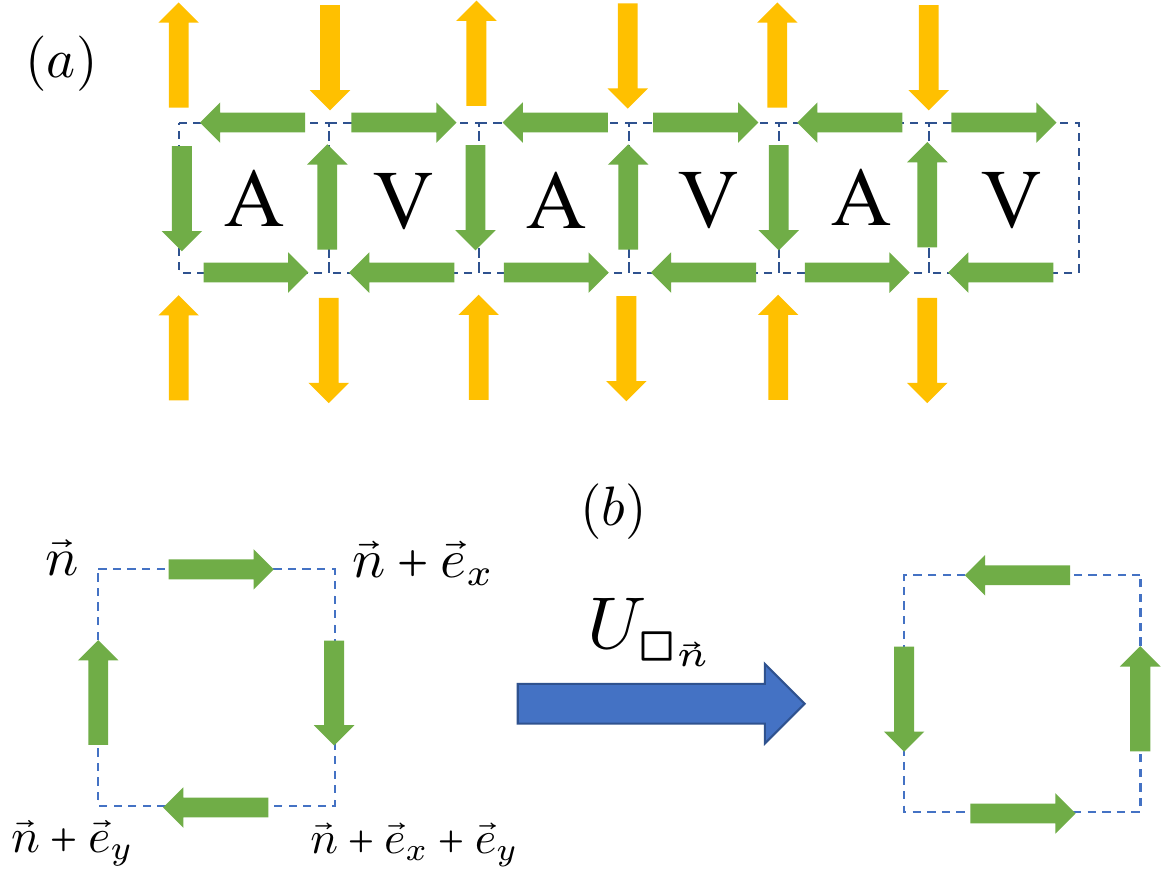}
\caption{(a) Schematic representation of the RK ladder. Green arrows indicate the spin-$1/2$ particles associated to each link of the ladder. Note that for the horizontal links we employ the notation $\rightarrow$~($\leftarrow$) 
to indicate spin $\uparrow$~($\downarrow$). The orange arrows denote virtual links, which we employ to define the local Gauss law at each lattice vertex~(see text).
The green spins are placed in the AVAVAV configuration, which we consider as the initial state in our simulations. (b) Application of the ring-exchange operator $U_{\square_{\vec n}}$ over the spins of plaquette $\square_{\vec n}$.}
\label{fig:1}
\end{figure}


 
 In this paper, we apply a quantum-classical approach, similar to that employed in Ref.~\cite{Klco2018}, for the study of the real-time dynamics of the RK model in a ladder geometry, possibly the simplest lattice gauge model with plaquette operators. We show that the combination of gauge invariance, symmetrization, and 
 the particular way plaquettes are blocked against ring-exchange in the ladder geometry, allow for an efficient mapping of the RK-ladder dynamics to a small number of qubits, an interesting feature for its implementation in NISQ devices. By employing gate scaling techniques, we show that the procedure allows for the successful simulation of the dynamics of small RK ladders in IBM-Q superconducting quantum computers. 


The structure of the paper is as follows. In Sec.~\ref{sec:Model} we introduce the RK-ladder model. Section~\ref{sec:mapping} is devoted to the mapping of the model into a set of effective equations. In Sec.~\ref{sec: QuantumSimulation} we discuss the actual simulation on a 
quantum computer. Our results are discussed in Sec.~\ref{sec:results}. Finally, we conclude in Sec.~\ref{sec:Conclusions}.
 


\section{Rohksar-Kivelson ladder}\label{sec:Model}

We consider a square ladder, as in Fig.~\ref{fig:1}~(a). 
We characterize each ladder vertex by a vector $\vec n=(n_x,n_y)$, with $n_y=1,2$ denoting the lower and upper legs, respectively. We associate to each link of the ladder 
a spin-$1/2$ degree of freedom, with two states $|\uparrow\rangle$ and 
$|\downarrow\rangle$. For convenience of the representation, for spins placed in links along the legs we denote those states as $|\rightarrow\rangle$ and $|\leftarrow\rangle$, respectively. Introducing the raising and lowering spin operators $S_{\vec n,\vec n'}^\pm$ acting on the spin placed at the link between the neighboring sites $\vec n$ and 
$\vec n'$~(
$S_{\vec n, \vec n'}^+ |\downarrow\rangle = |\uparrow\rangle$, 
$S_{\vec n, \vec n'}^- |\uparrow\rangle = |\downarrow\rangle$), we define for each plaquette $\square_{\vec n}$ (with $\vec n$ the top-left vertex of the plaquette, see Fig.~\ref{fig:1}~(b)) a ring-exchange operator 
\begin{eqnarray}
U_{\square_{\vec n}} &=& 
S_{\vec n, \vec n + \vec e_x}^- 
S_{\vec n + \vec e_x, \vec n + \vec e_x + \vec e_y}^+ 
S_{\vec n + \vec e_y, \vec n + \vec e_x + \vec e_y}^+ 
S_{\vec n, \vec n + \vec e_y}^-,
\label{eq:plaquette_operator}
\end{eqnarray}
with $\vec e_x=(1,0)$ and $\vec e_y=(0,1)$. Given a vortex~(antivortex) 
plaquette state $|V\rangle \equiv |\rightarrow,\downarrow,\leftarrow,\uparrow\rangle$~($|A\rangle \equiv |\leftarrow,\uparrow,\rightarrow,\downarrow\rangle$), where we follow the same clockwise order of Fig.~\ref{fig:1}~(b), the ring-exchange flips the state as $U_{\square_{\vec n}}|V\rangle = |A\rangle$, and  $U_{\square_{\vec n}}|A\rangle = |V\rangle$. Any other spin configuration, which we denote  below as $|B\rangle$, is blocked, i.e. non-flippable, and  $U_{\square_{\vec n}}|B\rangle=0$. 
We consider the RK model ~\cite{Rokhsar1988}:
\begin{equation}
H = \sum_{\square_{\vec n}} \left [-J( U_{\square_{\vec n}} + U_{\square_{\vec n}}^\dagger) + \lambda (U_{\square_{\vec n}} + U_{\square_{\vec n}}^\dagger)^2 \right ], 
\label{ham}
\end{equation}
where $\sum_{\square_{\vec n}}$ denotes the sum over all ladder plaquettes. Since $U_{\square_{\vec n}}^2|V\rangle = |V\rangle$,  
$U_{\square_{\vec n}}^2|A\rangle = |A\rangle$, 
$U_{\square_{\vec n}}^2|B\rangle = 0$, then the operator $\hat F = \sum_{\square_{\vec n}} (U_{\square_{\vec n}} + U_{\square_{\vec n}}^\dagger)^2$ acts as counter of the number of flippable, either $|V\rangle$ or $|A\rangle$, plaquettes.
The RK coupling $\lambda$ hence determines the energy penalty resulting from the change in the number of flippable plaquettes. We set below $J=1$. 

We define the local gauge transformation generator at vertex $\vec n$ as:
\begin{equation}
\hat G_{\vec n} = \hat S_{\vec n, \vec n + \vec e_x}^z - \hat S_{\vec n - \vec e_x,\vec n}^z +\hat S_{\vec n, \vec n + \vec e_y}^z - \hat S_{\vec n - \vec e_y, \vec e_n}^z,   
\end{equation}
with $\hat S_{\vec n, \vec n'}^z$ the $z$ projection of the spin placed in between two neighboring sites $\vec n$ and $\vec n'$. Note that in the ladder geometry, we have to consider virtual spins outside the ladder, which remain fixed during the dynamics~(orange arrows in Fig.~\ref{fig:1}). As a result, the gauge generators become of the form
\begin{eqnarray}
\hat G_{2n,1}&=&\hat S_{(2n,1),(2n+1,1)}^z - \hat S_{(2n-1,1),(2n,1)}^z \nonumber \\ 
&+& \hat S_{(2n,1),(2n,2)}^z+1/2,\\
\hat G_{2n+1,1}&=&\hat S_{(2n+1,1),(2n+2,1)}^z
-\hat S_{(2n,1),(2n+1,1)}^z \nonumber \\
&+&\hat S_{(2n+1,1),(2n+1,2)}^z-1/2, \\
\hat G_{2n,2}&=&\hat S_{(2n,2),(2n+1,2)}^z-\hat S_{(2n-1,2),(2n,2)}^z  \nonumber \\
&-&
\hat S_{(2n,1),(2n,2)}^z+1/2, \\
\hat G_{2n+1,2}&=&\hat S_{(2n+1,2),(2n+2,2)}^z
-\hat S_{(2n,2),(2n+1,2)}^z \nonumber \\
&-&\hat S_{(2n+1,1),(2n+1,2)}^z-1/2.
\end{eqnarray}
The RK-ladder Hamiltonian is then gauge invariant, $[\hat H, \hat G_{\vec n}]=0$, and thus the physical Hilbert space is also gauge invariant, i.e. it 
splits into sectors of eigenstates of $G_{\vec n}$. In the following, we focus 
on the charge-free sector, i.e. on states $|\psi\rangle$ that fulfill 
$G_{\vec n}|\psi\rangle = 0$, although a similar procedure as that discussed below can be applied to other sectors. We assume as well periodic boundary conditions along the ladder axis $x$.



\section{Mapping}\label{sec:mapping}
\begin{figure}[t!]
\includegraphics[width=0.9\columnwidth]{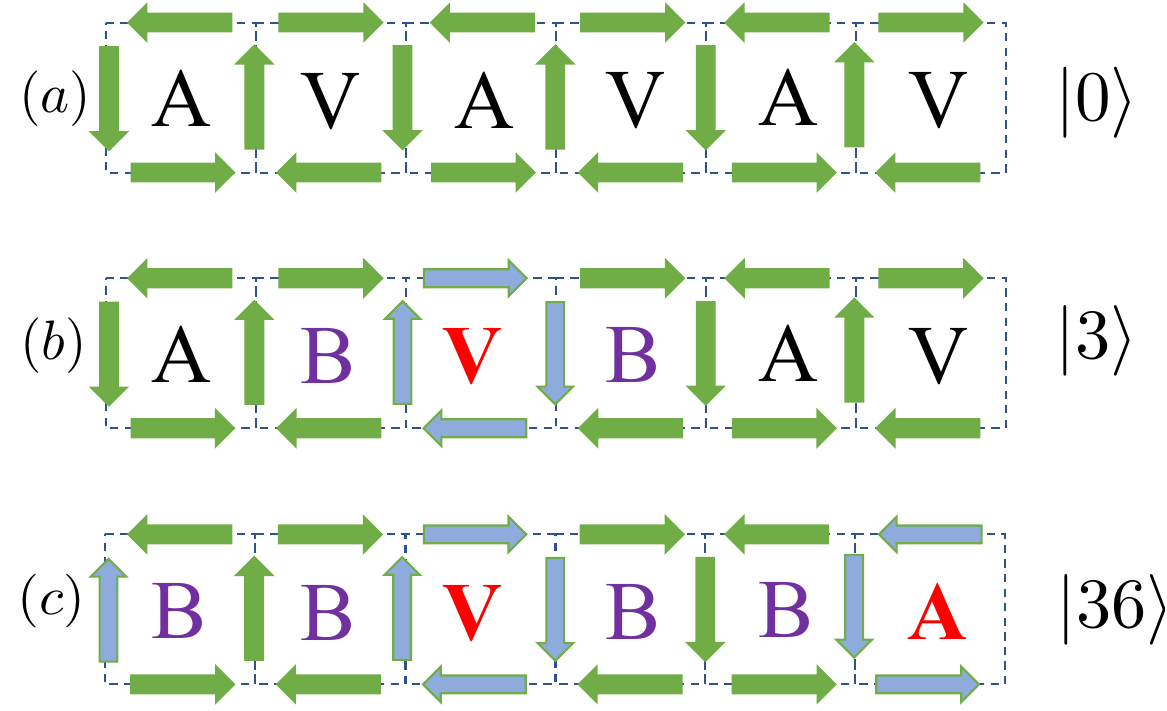}
\caption{(a) Initial plaquette configuration, $|0\rangle$. (b) The third plaquette is flipped, resulting in the state $|3\rangle$. (c) The third and the sixth plaquettes are flipped, leading to the state 
$|3,6\rangle$. Note that we do not need to determine explicitly the blocked plaquettes, or follow why they are blocked. The states are fully defined by stating which plaquettes have been flipped compared to the initial state (a).}
\label{fig:2}
\end{figure}

We are interested in the dynamics of the ladder plaquette, starting from a given initial configuration. A ladder with $N$ plaquettes~(and periodic boundary conditions) contains $3N$ spins, and hence presents in principle $2^{3N}$ spin configurations, which would hence naively require $3N$ qubits for its 
simulation. A more efficient procedure is possible, however, which we illustrate for the specific example of an RK-ladder with $6$ plaquettes initially prepared in an AVAVAV configuration. We denote this initial state as $\ket{0}$, which has obviously 
$6$ flippable plaquettes~(see Fig.~\ref{fig:2}~(a)). Other initial states are possible, which will modify the specific mapping discussed below, but similar mappings may be found as well. 



Flipping a plaquette blocks the two neighboring ones~(see Figs.~\ref{fig:2}~(b) and~(c)). For the mapping discussed below, it is crucial that in a ladder geometry, if 
flipping a plaquette blocks a neighboring one, the latter cannot be unblocked~(i.e.\
transformed back into a flippable plaquette) by flipping a third one. Due to this key property, and if we are only interested in the flippable or non-flippable character of the plaquettes, we do not need to track why a plaquette is blocked. This is crucially different in a two-dimensional~(2D) square lattice, where a blocked plaquette may be unblocked by flipping all the neighboring ones. As a result, in 2D geometries it is necessary to keep track why a plaquette is blocked, which results in a much more  resource-consuming algorithm. 

Due to the above-mentioned property, we may build the relevant states in a plaquette ladder by considering subsequent flips, without the need of describing the blocked plaquettes. We just need to keep track of which plaquettes have been flipped with respect to $|0\rangle$~(denoted in bold red in the examples of Fig.~\ref{fig:2}~(b) and~(c)), with the proviso that two neighboring plaquettes cannot be flipped. 
In the following, we employ the notation $\ket{ijk\dots}$, which denote ladders in which plaquettes $i,j,k,\dots$ have been flipped with respect to the initial state $\ket{0}$. Figures~\ref{fig:2}~(b) and~(c) show, respectively, the case of the states $|3\rangle$ and $|36\rangle$. Successive application of flips results in different families of states. States within a given family can be obtained from each other by translation taking into account periodic boundary conditions. 

For the case of $N=6$ plaquettes, the states with only one flip, $\ket{j}$, $j=1,\dots 6$, have four remaining flippable plaquettes, see the example of Fig.~\ref{fig:2}~(b). Clearly these states form a single family. States with two flips split into two families: $\{ \ket{13},\ket{15},\ket{24},\ket{26},\ket{35},\ket{46}\}$, with states with three flippable plaquettes,  
and $\{\ket{14},\ket{25},\ket{36}\}$, which have two flippable plaquettes. 
Finally, there are states with three flips $\{\ket{135}, \ket{246}\}$, which have three flippable plaquettes each.
Any other of the $2^{18}$ possible 
spin configurations of the ladder cannot be reached from $\ket{0}$, and hence does not need to be considered.
The relevant Hilbert space fragment contains 18 states. However, the number of states taking part in the dynamics is significantly smaller, since flips only link $|\psi_0\rangle \equiv |0\rangle$ to the symmetric superpositions of the states of each family:
\begin{eqnarray}
\ket{\psi_{1}} &=& \frac{1}{\sqrt{6}}( \ket{1} +\ket{2} +\ket{3} +\ket{4}+\ket{5} +\ket{6}),\\
\ket{\psi_{2}} &=& \frac{1}{\sqrt{6}} ( \ket{13} +\ket{15}+\ket{24} +\ket{26}+\ket{35} +\ket{46} ), \label{eq: 6plaq_basis}\\
\ket{\Bar{\psi_{2}}} &=& \frac{1}{\sqrt{3}} ( \ket{14} +\ket{25}+\ket{36} ),\\
\ket{\psi_{3}} &=& \frac{1}{\sqrt{2}} ( \ket{135} +\ket{246} ),
\end{eqnarray}
which form a closed set of states linked by the effective Hamiltonian $\hat{\mathcal{H}}_{eff}=-\hat{\mathcal{H}}_{eff}^{(0)} + \lambda\hat{\mathcal{H}}_{eff}^{(1)}$, where
\begin{eqnarray}
\hat{\mathcal{H}}_{eff}^{(0)} &=& 
\sqrt{6}|\psi_0\rangle\langle\psi_1| + 
2|\psi_1\rangle\langle\psi_2|  \nonumber \\
&+&
\sqrt{2} |\psi_1\rangle\langle\bar{\psi}_2|
+ \sqrt{3} |\bar{\psi}_2\rangle\langle\psi_3|
+\mathrm{H.c.} \\
\hat{\mathcal{H}}_{eff}^{(1)} &=& 
6|\psi_0\rangle\langle\psi_0| + 
4|\psi_1\rangle\langle\psi_1|+ 
3|\psi_2\rangle\langle\psi_2| \nonumber \\
&+& 2|\bar{\psi}_2\rangle\langle\bar{\psi}_2|+ 
3|\psi_3\rangle\langle\psi_3|.
\end{eqnarray}

Note that the large reduction in complexity, from $2^{18}$ basis states to just five effective states for the case of $6$-plaquette ladders, is based on the combination of gauge invariance, the above-mentioned property concerning blocked plaquettes, and symmetrization. This reduction of complexity allows for an efficient simulation of RK-ladders of a sizable number of plaquettes with a small number of qubits, well-suited for NISQ devices. 

Simulations of larger number of plaquettes demand the construction, using a simple classical numerical procedure, of the families of states with different number of flips that are equivalent when applying translation under periodic boundary conditions. Each family contributes to one state of the effective basis, given by the symmetrization of all the states in the family. The number of 
effective states determines the number of equations $N_{EQ}$, which demands the use of $Q=\log_2N_{EQ}$ qubits. For larger $N$, $Q\simeq 0.6N$, hence reducing the number of necessary qubits by a factor of $\simeq 5$ compared to the naive mapping of all spin states. For example, a sizable RK-ladder with $17$ plaquettes can be simulated, if noise is properly harnessed, with just $8$ qubits. 



\begin{figure*}[ht!]
  \centering
  \includegraphics[width=\textwidth]{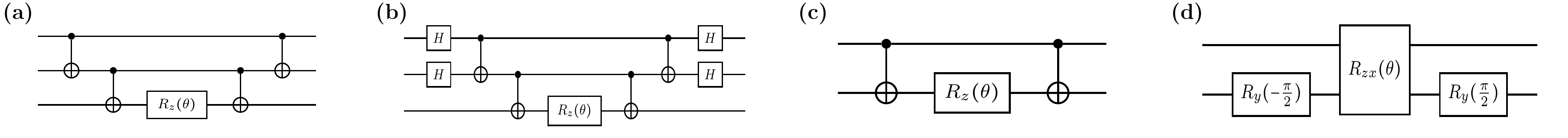}
  \caption{Standard circuit implementations of $(a)$ $e^{-iZZZ \delta t}$ and $(b)$ $e^{-iXXZ \delta t}$; $(c)$ $R_{ZZ}(\theta)$ and $(d)$ scaled $R_{ZZ}(\theta)$ based on cross-resonance formalism.}
  \label{fig:3}
\end{figure*}




\begin{table}[t!]
    \centering
    \begin{tabular}{|c|c|c|c|}
        \hline
        {\shortstack{Number of\\ plaquettes}} & {\shortstack{No. Pauli \\ terms}} & {\shortstack{\\No. CNOT \\(native)}} &  {\shortstack{No. CNOT\\ (scaled)}}  \\
        \hline
        4 & 7 & 6 & 0  \\
        \hline
        6 & 23 & 48 & 14  \\
        \hline
        8 & 26 & 64  & 20  \\
        \hline
    \end{tabular}
     \caption{For a given number of plaquettes, the table presents the number of Pauli terms obtained when decomposing the Hamiltonian into a sum of products of Pauli matrices, as in Eq.~\eqref{eq: H_eff}, the number of native~(IBM-Q) CNOT gates in the standard circuit implementation of unitary evolution of the Hamiltonian, and the number of remaining native~(IBM Q) CNOT gates in the scaled-gate circuit implementation of the unitary evolution operator.} 
     \label{table:1}
\end{table}



\section{Quantum Simulation}
\label{sec: QuantumSimulation}

We present below results for the dynamics of RK ladders with $4$, $6$, and $8$ plaquettes, which are described, respectively, by $3$, $5$ and $8$ effective states, and which may be then evaluated using, respectively, $2$, $3$, and $3$ qubits~(note that the simplest case with just two plaquettes, which we do not discuss, may be simulated with a single qubit). 
In order to simulate the dynamics with a quantum computer, we first decompose the effective Hamiltonian $\hat{\mathcal{H}}_{eff}$ into a 
string of products of Pauli matrices $X$ , $Y$ and $Z$, and the identity $I$~\cite{hantzko2023tensorized}. Table~\ref{table:1} provides the number of Pauli terms in the decomposition for different number $N$ of plaquettes in the RK ladder. For $N=6$, the $3$-qubit evaluation of the effective Hamiltonian in terms of Pauli terms acquires the form~(for $\lambda=1$):
\begin{eqnarray}
   \hat{\mathcal{H}}_{\text{eff}} &=& 
   2.25~ III - 0.612 ~IIX + 0.75 ~IIZ - 0.35~ IXI \nonumber \\
   &-& 0.5 ~IXX + 0.35~ IXZ - 0.5 ~IYY + 1.0~ IZI \nonumber \\
   &-& 0.612 ~IZ + 0.5 ~IZZ - 0.43~ XXI - 0.43 ~XXZ \nonumber \\
   &-& 0.43~ YYI - 0.43 ~YYZ + 1.5 ~ZII - 0.612~ ZIX \nonumber \\
   &-& 0.35 ~ZXI - 0.5 ~ZXX + 0.35 ~ZXZ - 0.5 ~ZYY \nonumber \\
   &+& 0.25 ~ZZI - 0.612~ ZZX - 0.25~ ZZZ. 
\label{eq: H_eff}
\end{eqnarray}

Expressing this linear superposition as 
$\hat{\mathcal{H}}_{\text{eff}}=
\sum_k \hat{h}_k$, we may then evaluate the evolution operator using first order Trotter-Suzuki approximation \cite{Trotter,Suzuki}
\eq{
e^{-i\hat{\mathcal{H}}_{\text{eff}}t} \approx \left ( \prod_k e^{-i\hat{h}_k\delta t} \right )^n
\label{eq: suzuki-trotter}
}
where $t=n \delta t$, $n$ is number of Trotter steps and $\delta t$ is the Trotter time step.
In a standard quantum circuit implementation,  each evolution operator $e^{-i\hat{h}_k \delta t}$ is decomposed
into single qubit rotations and CNOT gates \cite{nielsen_chuang_2010,quantgates}, as illustrated in the examples of Figs.~\ref{fig:3}~(a) and~(b). 
However, in NISQ devices it is crucial to reduce 
as much as possible the number and duration of two-qubit operations, which constitute the main source of 
errors. Alternative implementations have been hence recently proposed~\cite{Pulseefficient,Pekker,Chen}, in which the use of scaled quantum gates allows to decrease 
the number of two-qubit operations, leading to a 
significant error reduction in the implementation.

In IBM-Q machines, the CNOT operation is natively realized by the two-qubit rotation $R_{ZX}(\pi/2)$ where Pauli $Z$ and $X$ gates act on driven control and target qubits, respectively, implemented by echoed cross-resonance~(CR) pulses~\cite{CRgate}. A scaled $R_{ZX}(\theta)$ can be implemented with CR pulses, where the rotation angle $\theta$ depends on the pulse area. This scaled $R_{ZX}(\theta)$ is then used to implement a scaled $R_{ZZ}(\theta)$ as shown in Figs.~\ref{fig:3}~(c) and~(d). Note that $R_{ZZ}(\theta)$ is central in the implementation of any interaction term of the effective Hamiltonian. 
Since errors mostly arise due to the CR pulses, the use of scaled $R_{ZZ}(\theta)$ to implement all the terms of the evolution operator reduces the overall error in the circuit by decreasing both the overall duration of the pulse schedule and the number of native CNOT gates in the circuit. Table~\ref{table:1} provides the number of native CNOT gates in the basis gate set implementation, and the number of remaining CNOT gates in the scaled gate implementation for $N$ plaquettes in one circuit repetition. 
Appendix~\ref{appendix} presents a detailed description of the generation of scaled gates and of the analysis of the average gate fidelity for the scaled implementation of $R_{ZX}(\theta)$ against the standard CNOT implementation.




\begin{figure}[t]
  \includegraphics[scale=0.55]{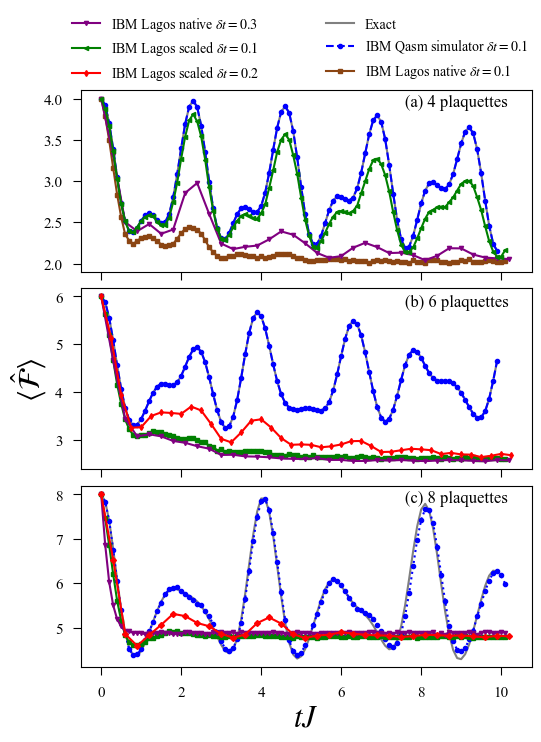}
  
  \caption{Average number of flippable plaquettes $\langle \hat F \rangle$ as a function of time for an RK-ladder with $\lambda=1$ and ~(a) $4$ ~(b) $6$ ~(c) $8$ plaquettes. We compare the results obtained from exact time evolution, the ideal simulator, and the noisy circuit with non-scaled and scaled gates for different Trotter steps $\delta t$~(see legend over Fig. (a)).}
  \label{fig:4}
\end{figure}



\begin{figure}[]
  \includegraphics[scale=0.55]{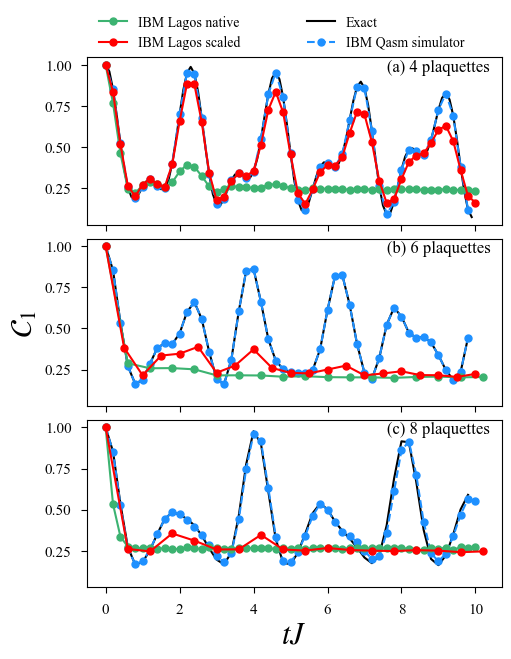}
  \caption{Plaquette-plaquette correlation  $\mathcal{C}_{1}$ as a function of time for an RK-ladder with $\lambda=1$ and ~(a) $4$ ~(b) $6$ ~(c) $8$ plaquettes. We compare the results obtained from exact time evolution, the ideal simulator, and the noisy circuit with non-scaled and scaled gates for Trotter step $\delta t= 0.3$ for non-scaled and $\delta t= 0.1$ for scaled gates.}
  \label{fig:5}
\end{figure}



\section{Results}
\label{sec:results}

In order to evaluate the dynamics and benchmark the results obtained using a quantum device, 
we monitor two observables that characterize the dynamics in the RK ladder: the average number of flippable plaquettes $\langle \hat F \rangle$, 
and the plaquette-plaquette correlations 
\begin{equation}
\mathcal{C}_{r} = 
\left \langle \left (U_{\square_{\vec n}} + U_{\square_{\vec n}}^\dagger \right )^2 \left (U_{\square_{\vec n + r\vec e_x}} + U_{\square_{\vec n + r\vec e_x}}^\dagger \right )^2\right \rangle.
\label{eq:Cij}
\end{equation}
Note that due to periodic boundary conditions, the correlation is independent of the particular choice of $\vec n$. 
The quantum simulation results discussed below were obtained using the noise model of the 7-qubit IBM-Q Lagos. In each simulation, we execute the quantum circuit for the unitary time evolution operator for 
Trotter steps with a given time step $\delta t$ and measure counts with 8192 shots per measurement. 

Figure~\ref{fig:4} shows our results for 
$\lambda=1$ and different number $N$ of plaquettes of  $\langle\hat{F}\rangle(t)$ up to a time $t=10/J$. 
The figure compares our results using exact-diagonalization~(ED, employing Krylov subspace), the ideal simulator, and the noisy circuit with non-scaled and scaled gates. 
As mentioned above, the minimal case of $N=2$ plaquettes can be solved with a single qubit, and as a result the two-plaquette quantum simulation in a noisy device matches exceptionally well with the ED calculations and ideal simulator calculations~(not shown). The advantage of using scaled quantum gates becomes evident when considering the case of $N=4$  plaquettes. The simulation for a noisy device with scaled gates with a Trotter step $\delta t= 0.1$ is in very good agreement with the ideal simulation, giving significant improvement over the noisy quantum simulation with the native gates as seen in Fig.~\ref{fig:4}(a), which clearly 
fails already at $t=0.5J$. 

For $N=6$ plaquettes, the number of CNOT gates is so large in the basis circuits that noisy quantum simulation with basis gates provides no useful data already for times $t > 0.7J$. In contrast, we see from Fig.~\ref{fig:4}(b) that the quantum simulation using scaled gates for a Trotter step $\delta t= 0.2$ recovers the qualitative behaviour of the ideal quantum simulation, which matches well with the ED calculation. For $N=8$ plaquettes the Trotter error remains very small in the ideal simulation. Again 
the noisy basis circuit simulation fails to recover useful data 
for any $t>0.2/J$, whereas the scaled gate simulation for a time step $\delta t= 0.2$ recovers the qualitative oscillation up to $t= 5J$, see Fig.~\ref{fig:4}(c).

Figure~\ref{fig:5} 
shows our results for the nearest neighbor correlation $C_{1}$ for RK-ladders with $N=4$, $6$ and $8$ plaquettes. We compared again the ED results, the ideal quantum simulation, and the noisy quantum simulation with non-scaled and scaled gates, using $\delta t=0.3$ for non-scaled gates and $\delta t=0.1$ for scaled gates. These results show again that 
for $N=4$ noisy quantum simulations with 
re-scaled gates result in an excellent quantitative agreement with ED calculations, whereas those with basis gates fail already at short times $t>2/J$. Moreover, we observe again that calculations with re-scaled gates provide a qualitatively correct behavior for $t<5/J$ for $N=6$ and $N=8$, whereas the calculations with basis gates
produce useless results already for $t>0.2/J$.



\section{Conclusions}
\label{sec:Conclusions}

We have shown that the dynamics of Rohksar-Kivelson ladders, 
a basic lattice gauge model, may be efficiently mapped into a small number of qubits due to the combination of gauge invariance, symmetrization, and the particular way plaquettes are blocked against ring-exchange. These properties reduce by a factor of $5$ the number of necessary qubits to simulate the dynamics of the RK ladder, an interesting feature for simulations in NISQ devices. We have illustrated the procedure using the simulator (including noise) of the 7-qubit IBM-Q machine Lagos, showing 
that the use of scaled gates allows for the faithful simulation of ladders of $N=4$ plaquettes. Moreover, we expect that using a combination of other error-mitigation techniques~\cite{Bharti2022}, in particular zero-noise extrapolation, should significantly improve the results for larger lattices, since even in this relative simple machine the qualitative behavior is well recovered at least up to $N\leq 8$ for up to $5$ ring-exchange times.


\acknowledgements
We acknowledge funding by the Volkswagen foundation and the Ministry of Science and Culture of Lower Saxony through \emph{Quantum Valley Lower Saxony Q1 (QVLS-Q1)}, and 
by the Deutsche Forschungsgemeinschaft (DFG, German Research Foundation) -- under Germany's Excellence Strategy -- EXC-2123 Quantum-Frontiers -- 390837967.

\bibliography{references}
\bibliographystyle{apsrev4-2}



\appendix

\onecolumngrid
\section{Generation of scaled gates}\label{appendix}

In this section, we discuss the generation of scaled $R_{ZZ}(\theta)$ gates from a scaled $R_{ZX}(\theta)$ gate using Qiskit Pulse~\cite{Pulse_2020,pulsesoftware}. Since a longer pulse duration results in more error, $R_{ZZ}(\theta)$ can be scaled to a lesser pulse schedule duration and hence lesser two-qubit error than the CNOT based implementation of $R_{ZZ}$. The scaled $R_{ZX}(\theta)$ is implemented by modifying the amplitude and duration of the $R_{ZX}(\pi/2)$ gate , which generates the fundamental entangling operation for a CNOT gate, as outlined in Ref.~\cite{Pekker,Chen}. The $R_{ZX}(\pi/2)$ is realized by echoed cross-resonance CR$(\pm\pi/4)$ pulses and an $X$-echoed $\pi$-pulse applied on the control qubit. The $X$-echoed $\pi$-pulse serves to minimize the effect of $ZI$ and $IX$ interaction terms~\cite{crosstalk,crham,Pulse_2020}. Effects of other terms like $ZZ$ and $IY$ can be suppressed by applying certain rotary pulses~\cite{rotary}. A perfect $R_{ZX}(\pi/2)$ gives rise to high fidelity entangling operation. The CR pulse has a Gaussian flat-top waveform, and has as attributes its flat-top width $w$, its amplitude $a$, its duration $d$, and the number of standard deviations of Gaussian tails contained in the pulse $n_{\sigma}$. The area $A$ under the Gaussian pulse is then given by 
\eq{
A = |a|w + |a|\sigma\sqrt{2\pi}\operatorname{erf}(n_{\sigma}).
}



\begin{figure}[h!]
  \includegraphics[scale=0.48]{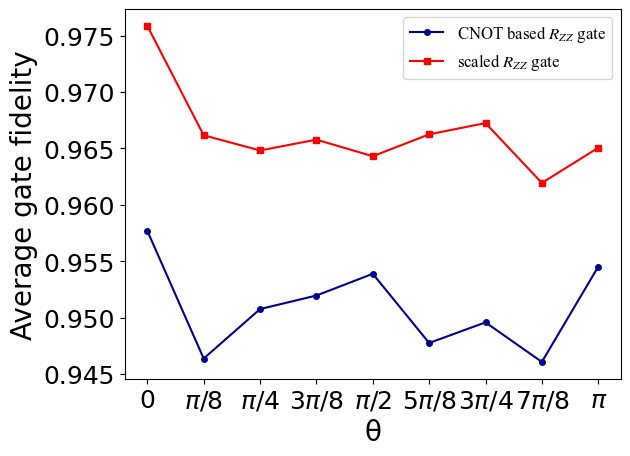}
  \caption{Comparison of the average gate fidelity for CNOT-based and scaled $R_{ZX}(\theta)$-based implementations of $R_{ZZ}$ for various angles, calculated for the IBM-Q Lagos device.}
  \label{fidelity analysis}
\end{figure}


To scale the pulse, depending on an arbitrary angle $\theta$, the area under the CR Gaussian pulse is modified relative to the area $A(\pi/2)$ of the CR pulse for $R_{ZX}(\pi/2)$ such that the area under the modified pulse $A(\theta)$ is given by
\eq{
A(\theta) = \frac{\theta}{\pi/2}A(\pi/2).
}
The modification in the pulse is achieved by either changing the width of the pulse or the amplitude, depending upon the initial parameters of the pulse. When $A(\theta) > |a(\pi/2)|\sigma\sqrt{2\pi}\operatorname{erf}(n_{\sigma})$, the pulse width is modified as 
\eq{
w(\theta) = \frac{a(\theta)}{|a(\pi/2)|} - \sigma\sqrt{2\pi}\operatorname{erf}(n_{\sigma}). 
}
When $A(\theta) < |a(\pi/2)|\sigma\sqrt{2\pi}\operatorname{erf}(n_{\sigma})$, the flat-top width becomes 0 and we scale the amplitude as
\eq{
|a(\theta)| = \frac{A(\theta)}{\sigma\sqrt{2\pi}\operatorname{erf}(n_{\sigma})}  
}

We perform a simple Quantum Process Tomography experiment using qiskit-experiments library~\cite{qiskitexp} to calculate the average gate fidelity for both the implementations of $R_{ZZ}$ for various angles. We find that the scaled-$R_{ZZ}$ has better fidelity than the CNOT-based implementation for all angles as shown in Fig.~\ref{fidelity analysis}.


\section{Mapping for ladders with $4$ and $8$ plaquettes}

Using the same notation as in the main text, the effective basis of states for the case of RK-ladders with $N=4$ plaquettes contains only $3$ states:
\eq{
\ket{\psi_{0}} &= \ket{0} \nonumber\\
\ket{\psi_{1}} &= \frac{1}{\sqrt{4}} ( \ket{1} 
+\ket{2} +\ket{3} +\ket{4} )\\
\ket{\psi_{2}} &= \frac{1}{\sqrt{2}} ( \ket{13} +\ket{24} )\nonumber
\label{basis4}
}
Using these basis states, we can rewrite the effective Hamiltonian as 
\begin{equation}
\hat{\mathcal{H}}_{eff}=
\begin{pmatrix}
 4\lambda & -2J & 0 \\
-2J & 2\lambda & -\sqrt{2}J \\
0 & -\sqrt{2}J & 2\lambda \\
\end{pmatrix}
  \label{eq: effective Hamiltonian 4}
\end{equation}
We map the Hamiltonian from $2^{12}$ states to an effective 3 states which can be simulated using 2 qubits.

For the case of $N=8$ plaquettes, the effective basis  has $8$ states:

We define the basis states as
\begin{align}
\ket{\psi_{0}} &= \ket{0} \nonumber\\
\ket{\psi_{1}} &= \frac{1}{\sqrt{8}}  (\ket{1} +\ket{2} +\ket{3} +\ket{4}+\ket{5} +\ket{6}+\ket{7} +\ket{8} )\nonumber\\
\ket{\psi_{2}} &= \frac{1}{\sqrt{8}} ( \ket{13} +\ket{17}+\ket{24} +\ket{28}+\ket{35} +\ket{46}+\ket{57} +\ket{68} ) \nonumber\\
\ket{\bar{\psi_{2}}} &= \frac{1}{\sqrt{8}} ( \ket{14} +\ket{16}+\ket{25}+\ket{27}+\ket{36}+\ket{38}+\ket{47}+\ket{58} )\nonumber\\
\ket{{\bar{\psi'_{2}}}} &= \frac{1}{\sqrt{4}} ( \ket{15} +\ket{26}+\ket{37}+\ket{48} ) \\
\ket{\psi_{3}} &= \frac{1}{\sqrt{8}} ( \ket{135} +\ket{137}+\ket{247} +\ket{248}+\ket{357}  +\ket{468}+\ket{157} +\ket{268} )\nonumber\\
\ket{\bar{\psi_{3}}} &=\frac{1}{\sqrt{8}} ( \ket{147} +\ket{146}+\ket{258}+\ket{257}+\ket{136}  +\ket{368}+\ket{247}+\ket{358} )\nonumber\\
\ket{\psi_{4}} &= \frac{1}{\sqrt{2}} ( \ket{1357} +\ket{2468}),\nonumber
\label{basis8}
\end{align}
Using these basis states, we can write the Hamiltonian in the form:
\begin{equation}
\hat{\mathcal{H}}_{eff}= \\
\\
\begin{pmatrix}
 8\lambda & -\sqrt{8}J & 0 & 0 & 0 & 0 & 0 & 0 \\
 \\
-\sqrt{8}J & 6\lambda & -2J& -2J &-\sqrt{2}J & 0 & 0 & 0\\
\\
0 & -2J & 5\lambda & 0 & 0 & -2J &-1J & 0 \\
\\
0 & -2J & 0 & 4\lambda& 0 & 0& -2J& 0\\
\\
0 &-\sqrt{2}J& 0& 0 &4\lambda & -\sqrt{2}J & 0 & 0  \\
\\
0 & 0&-2J &0& -\sqrt{2}J&4\lambda & 0& -\sqrt{4}J  \\
\\
0 & 0&-1J&-2J & 0&0 & 3\lambda& 0 \\
\\
0 & 0 & 0 & 0& 0&-\sqrt{4}J& 0 &4\lambda  \\

\end{pmatrix}
\\
\label{eq: effective Hamiltonian 8}
\end{equation}
%
%
The corresponding dynamics can be then simulated using only three qubits.

\end{document}